\documentclass{article}

\usepackage{amssymb, latexsym}
\usepackage{amsmath}
\usepackage{bm}
\usepackage{enumerate}
\usepackage{epsfig}

\begin{document}

\markboth{Nadia M. Abusag and David J. Chappell} {On sparse
reconstructions in near-field acoustic holography using the method
of superposition}

%
%

\title{ON SPARSE RECONSTRUCTIONS IN NEAR-FIELD ACOUSTIC HOLOGRAPHY USING THE METHOD OF SUPERPOSITION}

\author{NADIA M. ABUSAG and DAVID J. CHAPPELL\\ School of Science and Technology,\\ Nottingham Trent University,\\
Nottingham, UK}

\maketitle


\begin{abstract}
The method of superposition is proposed in combination with a sparse
$\ell_1$ optimisation algorithm with the aim of finding a sparse
basis to accurately reconstruct the structural vibrations of a
radiating object from a set of acoustic pressure values on a
conformal surface in the near-field. The nature of the
reconstructions generated by the method differs fundamentally from
those generated via standard Tikhonov regularisation in terms of the
level of sparsity in the distribution of charge strengths specifying the basis. In many cases, 
the $\ell_1$ optimisation leads to a solution basis whose size is
only a small fraction of the total number of measured data points.
The effects of changing the wavenumber, the internal source surface
and the (noisy) acoustic pressure data in general will all be
studied with reference to a numerical study on a cuboid of similar
dimensions to a typical loudspeaker cabinet. The development of
sparse and accurate reconstructions has a number of advantageous
consequences including improved reconstructions from reduced data
sets, the enhancement of numerical solution methods and wider
applications in source identification problems.
\end{abstract}


\section{Introduction}

Near-field acoustic holography (NAH) was first documented in $1980$
\cite{EW80a,EW80b} as a method for reconstructing acoustic radiation
from vibrating structures based on acoustic pressures measured in a
hologram plane. This was originally carried out using Fourier
acoustics and so was only suitable for geometries corresponding to
the separable geometries of the acoustic wave equation such as an
infinite plane, an infinite cylinder or a sphere in a free field.
Recent progress \cite{GC12} has seen ideas from signal processing
and imaging, and in particular the much celebrated compressive
sampling (or compressed sensing) \cite{EC08}, adopted to reduce
measurement requirements and reconstruct acoustic fields above the
Nyquist frequency. In this paper we develop a method based on an
inverse formulation of the superposition method proposed by Koopman
\emph{et al.} \cite{GK89} that can reconstruct acoustic fields from
arbitrary shaped radiating objects and make use of sparse
reconstruction principles as proposed in Ref. \cite{GC12}.

Near-field acoustic holography for arbitrary geometries was first
performed using the inverse boundary element method (IBEM)
\cite{BG88}. A large number of publications on IBEM have since
emerged, and more recently with the IBEM combined in hybrid methods
which employ expansions of particular solutions of the Helmholtz
equation to enrich the measured pressure data \cite{SW02,SW04}. One
drawback of the IBEM is that it suffers from the same irregular
frequency problem (at the resonances of the associated interior
domain) as the forward boundary element method (BEM) \cite{DC09}.
Whilst this problem can be treated using the same methods as for the
forward problem (for example, using the Burton and Miller method
\cite{AB71} as in Ref. \cite{DC09}, or the so-called combined
Helmholtz integral equation formulation (CHIEF) \cite{HS68}), such
methods complicate the implementation and reduce the computational
efficiency to some degree.

In addition to the irregular frequency problem, boundary element
approaches also require very fine meshes to accurately represent
solutions at high frequencies, and to give good representations of
the acoustic field in the immediate vicinity of the radiating
object. A number of methods have been suggested to resolve the
shortcomings of IBEM in a manner that is both computationally
efficient and relatively straightforward to implement. These methods
effectively fit the measured pressure data to a linear combination
of basis functions, and then use the coefficients determined through
this approximation to determine the normal velocity of the vibrating
object. The most well-known of these methods with respect to NAH are
the Helmholtz equation least squares (HELS) method proposed by Sean
Wu and co-workers \cite{ZW97,SW00} and the method of superposition,
which was applied to NAH in Refs. \cite{AS05,NV06}, including a
comparison against boundary element approaches in the latter case.
For HELS, the basis functions are chosen as particular solutions to
the Helmholtz equation on an idealised domain, typically a sphere.
In the case of the superposition method then it is the free space
Green's functions that are employed for the basis.

It is shown in Ref. \cite{GK89} that the superposition method for an
exterior acoustic problem is equivalent to the Helmholtz integral
equation. This is one reason for favoring the superposition approach
to HELS, where the chosen basis is typically only complete outside
the minimum sphere enclosing the radiating object \cite{SW00}. This
would be a drawback for approximating radiation from objects that
are far from spherical in the region inside this minimum sphere.
Koopman \emph{et al.} \cite{GK89} also argue that the superposition
method will not suffer from the same irregular frequency problem as
the boundary element method since the set of source points chosen
for the superposition (after truncating the superposition integral
to a finite sum) will not form a unique boundary surface inside the
interior volume. However, the computations in Ref. \cite{GK89} were
only performed with small numbers of source points, and irregular
frequency problems may arise should the source points more closely
represent an interior boundary surface \cite{WB97,AL10}.

One of the main challenges involved in the superposition method is
obtaining an optimal choice of source (or charge) points over which
to compute the superposition. In particular, research on (forward)
interior Helmholtz problems \cite{AB08} suggests that the source
points should be chosen close to the radiating surface so that no
singularities of the analytically continued solution lie between the
charge points and the radiating surface. However, for the exterior
problem, Koopman \emph{et al.} \cite{GK89} report that choosing the
charge points too close to the radiating surface degrades the
accuracy of the method, whilst choosing charge points too far from
the surface leads to very poor conditioning. Applying the
superposition method together with a nonlinear optimisation
algorithm to optimise the accuracy of the solution over both the
charge point strengths and locations is known as the method of
fundamental solutions, see for example Refs. \cite{GF98,GF03}. We
will show that in combination with sparse optimisation methods, a
relatively small number of charge points can be chosen to accurately
reconstruct the surface velocity, particularly in the sub-Nyquist
frequency regime.

Regularisation schemes for NAH are typically based on standard
Tikhonov methods, an excellent overview is presented in Ref.
\cite{EW01}. The most favorable method for selecting the
regularisation parameter in NAH applications is reported as a
modified generalised cross validation (GCV) method, although many
other possibilities are discussed in Ref. \cite{EW01}. Recent work
\cite{GC12} has suggested the possibility of using $\ell_1$
optimisation algorithms to instead in cases when a suitable
dictionary of basis functions can be defined. Whilst this leads to a
loss of efficiency in the optimisation process, it also promotes
sparsity of the solution, and the convexity of the $\ell_1$ norm
means that convex optimisation toolboxes such as SPGL1 \cite{EB15}
or CVX \cite{MG15} can be applied. Chardon \emph{et al.} \cite{GC12}
combine sparse optimisation techniques with a Fourier basis / plane
wave approximation of the solution on flat star-like plates and
report accurate reconstructions with sparse basis expansions. In
combination with randomised measurement data, these methods gave
faithful reconstructions, even above the Nyquist frequency. More
recently, sparse $\ell_1$ optimisation with a plane wave basis has
been proposed as a means of source identification from spherical
acoustic radiators \cite{AX15a}.

In this paper we discuss the possibility of a sparse solution
representation using $\ell_1$ optimisation techniques based on the
superposition method for application to general three dimensional
problems. A study of this combination of methods for reconstructing
vibration of flat plates has recently been presented in Ref.
\cite{AX15b}. The method of superposition appears to be the most
promising approach since there are close links between plane wave
method and the method of superposition for bounded domains as
described in Ref. \cite{CA05}. In addition, the superposition method
is typically accurate for frequencies above the six degrees of
freedom per wavelength rule of thumb for boundary element methods
\cite{AB08} and can handle internal resonance frequencies in a
relatively simple manner \cite{WB97,AL10}. The application and
suitability of these methods will also be considered for the
reconstruction of the Neumann boundary data on a cuboid of similar
dimensions to a typical loudspeaker cabinet over a range of
frequencies.

Initially we will study test problems where the exterior acoustic
field is generated by a monopole point source inside the structure.
In this case the solution could be represented exactly for
arbitrarily high frequencies using the method of superposition if
one of the points on the surface of interior charge points coincides
with the monopole generating the acoustic field. However, in general
such information would not be known \emph{a-priori} when sampling an
acoustic field in NAH. We will therefore examine the dependence of
the reconstruction on the locations of the charge points and the
monopole generating the acoustic field. Finally, we will consider a
case where the acoustic field is not generated by an interior
monopole, but rather by a relatively small vibrating patch on an
otherwise rigid structure. This leads to a problem where the method
of superposition should return a sparse expansion for interior
charge point surfaces very close to the physical boundary, since the
weighting of sources close to the vibrating region will be dominant
compared to those close to the rigid part of the structure. The
latter (locally radiating) case also has parallels with the
situation that arises when modelling acoustic radiation from
loudspeakers.

\section{Method of Superposition for Near-field Acoustic Holography}

Let $\Omega\subset\mathbb{R}^{3}$ be a finite domain with boundary
surface $\Gamma$. Let
$\Omega_{+}=\mathbb{R}^{3}\setminus\bar{\Omega}$ denote the
unbounded exterior domain, which is assumed to be filled with a
homogeneous compressible acoustic medium with density $\rho$ and
speed of sound $c$. For a time-harmonic disturbance of frequency
$\omega$, the sound pressure $p$ satisfies the homogeneous Helmholtz
equation in $\Omega_{+}$
\begin{equation} \label{HE1}
\Delta p+k^{2}p=0,
\end{equation}
where $k=\omega/c$ is the wave-number. Since this work considers an
unbounded exterior domain, then $p$ must also satisfy the Sommerfeld
radiation condition. 
The superposition method approximates $p$ at some point
$\mathbf{x}\in\bar{\Omega}_+$ using a basis expansion of the form
\begin{equation}\label{MFS1}
p(\mathbf{x})\approx\sum_{j=1}^{n} \sigma_j
G_{k}(\mathbf{x},\mathbf{y}_j),
\end{equation}
where $G_{k}$ is the free space Green's function for Helmholtz
equation in three dimensions given by
\begin{equation}\label{GF1}
G_{k}(\mathbf{x},\mathbf{y})=\frac{e^{ik|\mathbf{x}-\mathbf{y}|}}{4\pi
|\mathbf{x}-\mathbf{y}|}.
\end{equation}
Here $\mathbf{y}_i\in\Omega$, $i=1,...,n$ are the source locations
and $\sigma_i$ are the source strengths, which are determined by
application of the method.


In the NAH problem we are given values of the acoustic pressure $p$
at a discrete set of points the acoustic near field within
$\Omega_{+}$. We will assume that the data points $\mathbf{x}_i$,
$i=1,...,m$ lie on a surface $\Gamma^{*}\subset\Omega_{+}$. Note
that the pressure data is usually obtained from measurements using a
microphone array. However, in this work we only generate the
pressure data numerically as described in Section \ref{sec:num}. The
NAH problem is to use the given pressure data to recover the Neumann
boundary data on $\Gamma$. Solving this problem via the method of
superposition is then a matter of finding the set of source
strengths $\sigma_j$, $j=1,...,n$, that reproduce the acoustic
pressure data to some desired accuracy in the least squares sense.
That is, $\sigma_j$ are chosen so that the $\ell_2$ norm of the
residual vector $\mathbf{r}$, with entries given by
\begin{equation}\label{MFSmin}
r_i=p(\mathbf{x}_i)-\sum_{j=1}^{n} \sigma_j
G_{k}(\mathbf{x}_i,\mathbf{y}_j)
\end{equation}
for $i=1,..,m$, is smaller than a desired error tolerance. Once the
source strengths have been obtained then the Neumann boundary data
can be recovered from
\begin{equation}\label{MFS2}
\frac{\partial
p}{\partial\mathbf{n}}(\mathbf{x})\approx\sum_{j=1}^{n} \sigma_j
\frac{\partial G_{k}}{\partial\mathbf{n}}(\mathbf{x},\mathbf{y}_j),
\end{equation}
where $\mathbf{n}$ is the outward unit normal to $\Gamma$. Note that
the linearised Euler equation for time harmonic waves leads to the
following simple relationship between the Neumann boundary data
computed here, and the normal velocity $v$ of the radiating object
\begin{equation}\label{nv}
\frac{\partial p}{\partial\mathbf{n}}=i\omega\rho v.
\end{equation}
Regularisation is always required in general, even for $n=m$, since
NAH is an ill-posed inverse problem (see for example \cite{DC09}).
For experimental problems, the pressure measurements will contain
errors and the ill-posedness of the problem means that these errors
are amplified in the (unregularised) solutions, often rendering them
meaningless. Most previous work on NAH has concentrated on using
Tikhonov regularisation, together with generalised cross validation
(GCV). In the next section we describe a scheme designed to promote
sparsity in the solution as suggested in Ref. \cite{GC12} for
two-dimensional planar problems.

\section{Regularisation and Sparse Reconstruction}

One of the oldest and most widely used regularisation methods is
Tikhonov regularisation, which applied to equation (\ref{MFSmin})
leads to the following minimisation problem
\begin{equation}\label{HE4reg}
\hat{\boldsymbol{\sigma}}=\mathrm{arg}\min_{\boldsymbol{\sigma}}\left\{\|\mathbf{r}\|_2^2+\lambda\|L\boldsymbol{\sigma}\|_2^2\right\},
\end{equation}
where $L$ is the Tikhonov matrix (most simply chosen as the identity
matrix) and $\lambda>0$ is a regularisation parameter to be
determined. We have also introduced the notation
$\boldsymbol{\sigma}=[\sigma_1\:\:\sigma_2\:\ldots\:\sigma_n\:]^T$
for the vector containing the source strengths, and likewise
$\hat{\boldsymbol{\sigma}}$ is the vector containing the regularised
and reconstructed source strengths at the $n$ interior source
points.

In this work we adopt the alternative regularisation approach of
Chardon \emph{et al.} \cite{GC12}, which favors sparse
representations of the solution. In other words, it (approximately)
minimises $|\boldsymbol{\sigma}|_0$, the number of non-zero entries
of $\boldsymbol{\sigma}$. As noted by Chardon \emph{et al.}
\cite{GC12}, the possibility of a sparse reconstruction is highly
dependent on the basis functions used to represent the solution. In
the superposition method, these basis functions are the fundamental
solution of the Helmholtz equation at a set of distinct interior
charge points. The results in Koopman \emph{et al.} \cite{GK89}
suggest the feasibility of sparse solution representations for a
large range of wavenumbers using the superposition method. The
results presented in the next section will investigate the
conditions whereby high quality sparse representations are indeed
possible. For the examples considered here we expect that the
quality of the solutions attainable will be highly dependent on the
location of the interior charge points used for the superposition.
The acoustic pressure data is created by either a single interior
monopole, or by applying the BEM to give the acoustic field radiated
from a relatively small vibrating patch. We will investigate whether
the sparse reconstruction approach can pick out solutions
$\boldsymbol{\sigma}$ that make use of the underlying sparsity in
these examples, where this sparsity arises either due to the low
number of monopoles needed to generate the field in the former case,
or due to the relatively small region over which the reconstructed
field is dominant in the latter case.

Directly minimising  $|\boldsymbol{\sigma}|_0$ is often intractable
due to non-convexity \cite{GC12}. We therefore instead seek to
minimise the $\ell_1$ norm
\begin{equation}
\|\boldsymbol{\sigma}\|_1=\sum_{j}|\sigma_j|.
\end{equation}
The use of the $\ell_1$ norm allows one to apply powerful convex
optimisation algorithms and still promotes sparsity by making many
of the components of $\boldsymbol{\sigma}$ negligibly small, meaning
that they can be well approximated by zero without degrading the
reconstructed solution.  The following procedure will be applied to
find a sparse representation $\hat{\boldsymbol{\sigma}}$ of the
source strengths $\boldsymbol{\sigma}$
\begin{equation}\label{L1reg1}
\hat{\boldsymbol{\sigma}}=\mathrm{arg}\min_{\boldsymbol{\sigma}}\|\boldsymbol{\sigma}\|_1\:\:\:\mathrm{subject}\:\:\mathrm{to}\:\:\:\|\mathbf{r}\|_{2}^2\leq\epsilon.
\end{equation}
This procedure will be implemented using the convex optimisation
toolbox CVX \cite{MG15}. This procedure requires a data fidelity
constraint $\epsilon$ to be specified. Choosing this parameter
involves a trade off between allowing sparser solutions with larger
values of $\epsilon$ and achieving more accurately reconstructed
solutions with smaller values of $\epsilon$. Chardon \emph{et al.}
\cite{GC12} recommend a choice of $\epsilon$ of the order $20\%$ to
$30\%$ of the $\ell_2$ norm of the measured pressure data. However,
a good choice of $\epsilon$ is likely to depend on how noisy the
pressure data is and hence will be problem dependent.

For completeness, and to emphasise the links between the $\ell_1$
regularisation approach and Tikhonov regularisation we note that
(\ref{L1reg1}) may be expressed in the form \cite{GC12,SC99}
\begin{equation}\label{L1reg2}
\hat{\boldsymbol{\sigma}}=\mathrm{arg}\min_{\boldsymbol{\sigma}}\left\{\|\mathbf{r}\|_2^2+\lambda\|\boldsymbol{\sigma}\|_1\right\}.
\end{equation}
This procedure is known as the Basis Pursuit Denoising (BPDN) and is
introduced in Sect. 5.1 of Ref. \cite{SC99}, where the interested
reader can find further details, including a discussion of suitable
choices of $\lambda$ in the presence of standard Gaussian noise.
Here we simply note the parallels between the expression
(\ref{L1reg2}) and equation (\ref{HE4reg}), and in particular that
one of the main differences is the norm employed in the final term.
In particular, the $\ell_1$ norm replaces the square of the $\ell_2$
norm in the Tikhonov case, and it is this difference that promotes
sparsity in the $\ell_1$ approach. The differences between these two
approaches will be investigated numerically in the next section.

\section{Numerical Results}\label{sec:num}

Numerical results will be computed for acoustic radiation from a
cuboid of similar dimensions to a typical loudspeaker cabinet
($0.28\mathrm{m}\times0.28\mathrm{m}\times0.42\mathrm{m}$). Although
the method of superposition is a mesh free method, we will use a
triangulation of the cuboid to generate the points at which the
pressure data is computed, as well as the internal charge points and
the points at which we reconstruct the solution on $\Gamma$. In
particular, for a given triangulation of $\Gamma$ we reconstruct the
Neumann boundary data at the centroid of each triangle and project
(from each centroid) a distance $\delta$ along the normal vector to
$\Gamma$ into $\Omega_+$ to obtain the points where the exterior
pressure data is recorded. The internal charge points are positioned
on a cuboid inside $\Omega$, which is just a scaled down version of
$\Gamma$ with scaling factor $\alpha\in(0,1)$. For example, a value
of $\alpha=0.5$ corresponds to a cuboid surface of internal charge
points whose dimensions are exactly half those of $\Gamma$. Assuming
that $\Omega$ is centred at the origin, we simply take a point
$\mathbf{x}\in\Gamma$ and multiply by $\alpha$ to obtain the
corresponding point $\mathbf{y}$ on the internal surface of charge
points thus
\begin{equation}\label{alpha}
\mathbf{y}=\alpha\mathbf{x}.
\end{equation}

Initially we reconstruct the boundary data generated by a point
source at $\mathbf{x}_{0}=(0,\:0,\:z_{0})\in\Omega$, where $\Omega$
is centred at the origin. The pressure data is therefore of the form
\begin{equation}\label{ps1}
(\mathbf{p}_{0})_{j}=a\frac{e^{ik|\mathbf{x}_{0}-\mathbf{x}_{j}|}}{|\mathbf{x}_{0}-\mathbf{x}_{j}|},\hspace{1cm}j=1,...,m,
\end{equation}
where $a\in\mathbb{C}$ is the strength of the source, which in these
examples is arbitrarily taken to be $a=3-i$. The boundary data
generated at $\mathbf{y}\in\Gamma$ may also be obtained for the case
of a point source at $\mathbf{x}_{0}=(0,\:0,\:z_{0})$ by replacing
$\mathbf{x}_{j}$ in (\ref{ps1}) by $\mathbf{y}\in\Gamma$,
differentiating in the direction of $\mathbf{n}_{y}$ and evaluating
at the centroids of the triangulation $\mathbf{y}=\mathbf{y}_{j}$
for $j=1,...,m$ to give
\begin{equation}\label{ps2}
(\mathbf{v})_{j}=a\left(\frac{\mathbf{n}_{y_{j}}\cdot(\mathbf{x}_{0}-\mathbf{y}_{j})}{|\mathbf{x}_{0}-\mathbf{y}_{j}|^{3}}(1-ik|\mathbf{x}_{0}-\mathbf{y}_{j}|)e^{ik|\mathbf{x}_{0}-\mathbf{y}_{j}|}\right),\hspace{1cm}j=1,...,m.
\end{equation}
Using this calculation it is possible to verify the accuracy of the
(Tikhonov or $\ell_1$) regularised approximate solutions with
different wavenumbers and singularity point positions $z_{0}$. We
will also investigate the behavior of the method at irregular
frequencies of the volume enclosed by the interior charge points,
and the dependence on the dimensions / location of the interior
charge point surface controlled by the parameter $\alpha$.

Uniformly distributed and additive white noise will be applied to
$\mathbf{p}_{0}$ in order to more closely replicate experimental
observations. The use of Gaussian noise was also considered and, in
general, led to slightly more accurate reconstructions than
uniformly distributed noise. However, the quality of the
reconstructions also fluctuated more widely when using different
Gaussian noise vectors (of the same norm) than for uniformly
distributed noise, and so we present the results for uniformly
distributed noise since we believe they give a more indicative and
repeatable measure of the performance of our reconstruction methods.
We denote the added noise vector as $\mathbf{w}$ and specify the
ratio
\begin{equation}\label{SNR}
w=\frac{\|\mathbf{w}\|_2}{\|\mathbf{p}_{0}\|_2},
\end{equation}
referring to $w$ as the level of added noise in the sequel. Note
that $w$ is related to the standard signal to noise ratio (SNR) via
$\mathrm{SNR}=w^{-2}$, or in decibels,
$\mathrm{SNR}_{\mathrm{dB}}=10\log_{10}\left(w^{-2}\right)$.

Finally, we consider the case of reconstructing a locally vibrating
structure, where only a small region of the structure is vibrating.
Such an assumption is typical for the case of a loudspeaker and also
allows us to study a case where the locations of any singularities
of the continuation of the acoustic field into $\Omega$ are unknown,
as is usually the case in practice. Here the acoustic pressure data
will be generated using the boundary element method applied to the
forward Neumann problem described earlier. Through this example we
demonstrate the broader applicability of the sparse reconstruction
algorithm, where the structure of the measured signal should have a
sparse structure, but not necessarily the basis for the
superposition method.

\subsection{Comparison with Tikhonov regularisation}

First consider the case $k=1$ and $z_{0}=0.05$, where the frequency
is relatively low, is not close to an irregular frequency and
$\mathbf{x}_{0}$ is relatively close to the origin and will lie
inside the surface on which the interior charge points are located.
Under such conditions the superposition method is expected to work
well. Table \ref{T1} shows the $\ell_2$ percentage errors in the
reconstructed Neumann boundary data for three different
regularisation strategies and differing noise levels. The three
solution strategies to be compared are (i) $\ell_1$ regularisation
and taking the sum over all interior charge points, (ii) sparse
$\ell_1$ regularisation where only contributions from dominant
charge points are considered, and (iii) standard Tikhonov
regularisation using Generalised Cross Validation (GCV) to determine
the regularisation parameter. In the latter case the computations
have been performed using Hansen's extensive regularisation toolbox
for Matlab \cite{PH07}. In the case of the sparse reconstruction,
the criteria used to determine whether the $j$th charge point is
dominant is if
$$\log\left(\frac{|\sigma_j|}{\min_i|\sigma_i|}\right)>\beta\log\left(\frac{\max_i|\sigma_i|}{\min_i|\sigma_i|}\right).$$
We will use the notation $N^*(\beta)$ for the number of dominant
charge points satisfying this condition, taking $\beta=0.5$ by
default and so we denote $N^*=N^*(0.5)$. The $\ell_2$ percentage
error in the reconstructed solution $\hat{\mathbf{v}}$ is calculated
using
\begin{equation}\label{l2err}
\frac{\|\hat{\mathbf{v}}-\mathbf{v}\|_2}{\|\mathbf{v}\|_{2}}\times100\%.
\end{equation}

The pressure data are specified at a distance $\delta=0.035$m from
$\Gamma$ and the internal source surface is scaled down to have
dimensions $\alpha=1/3$ the size of $\Gamma$. We note that these
choices should lead to good results based on the fact that $\delta$
should be chosen small enough to capture evanescent contributions to
the pressure field, but still large enough to be a practical
distance for taking experimental measurements. For a further
discussion on the role of evanescent wave contributions in
superposition methods for NAH problems, the interested reader is
referred to Sect. 5 of Ref. \cite{NV06}. For the choice of the
parameter $\alpha$ (\ref{alpha}), Koopman \emph{et al.} \cite{GK89}
suggest that too small a value will lead to severe ill conditioning
as the charge points become very close together, but choosing too
large a value of $\alpha$ will also give poor results. A choice in
the range $\alpha\in(0.1,0.6)$ is advised in Ref. \cite{GK89}. It is
also beneficial for the charge point surface to enclose any
singularities of the associated interior problem \cite{AB08}. For
these experiments the number of charge points, the number of
measurement points and the number of points at which we reconstruct
the solution are all equal to 576. This is achieved by triangulating
the internal source surface in an identical way to $\Gamma$ and
taking the charge points at the triangle centroids.

The results in Table \ref{T1} show that in the noise free case, the
reconstruction errors for both the full $\ell_1$ method and Tikhonov
regularisation are very small with Tikhonov reconstruction
performing better. A sparse representation of the solution is not
feasible here in general unless one of the charge points coincides
with the monopole generating the acoustic field; the $\ell_1$
optimisation identifies a relatively large number $N^*=98$ of
dominant sources and the error of the `sparse' reconstruction
increases significantly compared with the reconstruction using all
$576$ source points. However, once noise is present in the pressure
data then sparse representations of the solution can be obtained
with a similar level of accuracy to the Tikhonov approach. The
reason for this can be explained by considering how the data
fidelity parameter $\epsilon$ is chosen in (\ref{L1reg1}). In
particular we take
\begin{equation}\label{datfid}
\epsilon=(\max\{\epsilon_{\text{min}},w\})^2\|\mathbf{p}_{0}\|_2^2,
\end{equation}
where $w$ is the level of noise added to the pressure data as
before. Recall that larger choices of $\epsilon$ permit sparser
solution representations. However, it only makes sense to choose a
larger $\epsilon$ for noisy data, otherwise it leads to less
accurate reconstructions. The parameter $\epsilon_{\text{min}}\geq0$
is included as a tolerance level that is used for the low or zero
noise case. A relatively large choice of $\epsilon_{\text{min}}$
will lead to sparser reconstructions at the expense of accuracy, and
the converse is true for small $\epsilon_{\text{min}}$. The results
in this work have been obtained with
$\epsilon_{\text{min}}=\textsc{1e-6}$.

\begin{figure}[h]
\begin{center}
          \epsfig{file=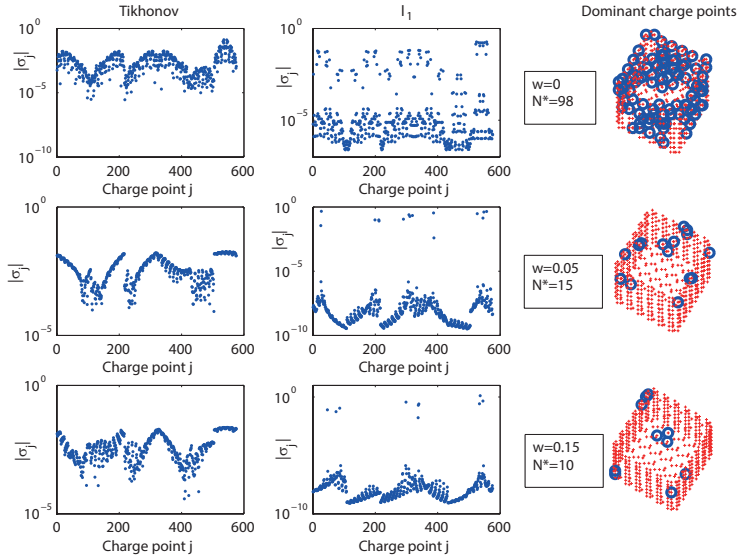, width=\textwidth}
        \caption{\label{azzplot} Left and centre columns: A comparison of the charge point strengths using both the Tikhonov and $\ell_1$ approaches for wavenumber $k=1$ and exterior pressure data generated by a point source at $(0,0,0.05)$. Right column: the locations of the dominant charge points for the $\ell_1$ approach. The top row shows the case of clean pressure data and the other rows show the results for differing levels of added noise $w$.}
\end{center}
\end{figure}

\begin{figure}[h]
\begin{center}
          \epsfig{file=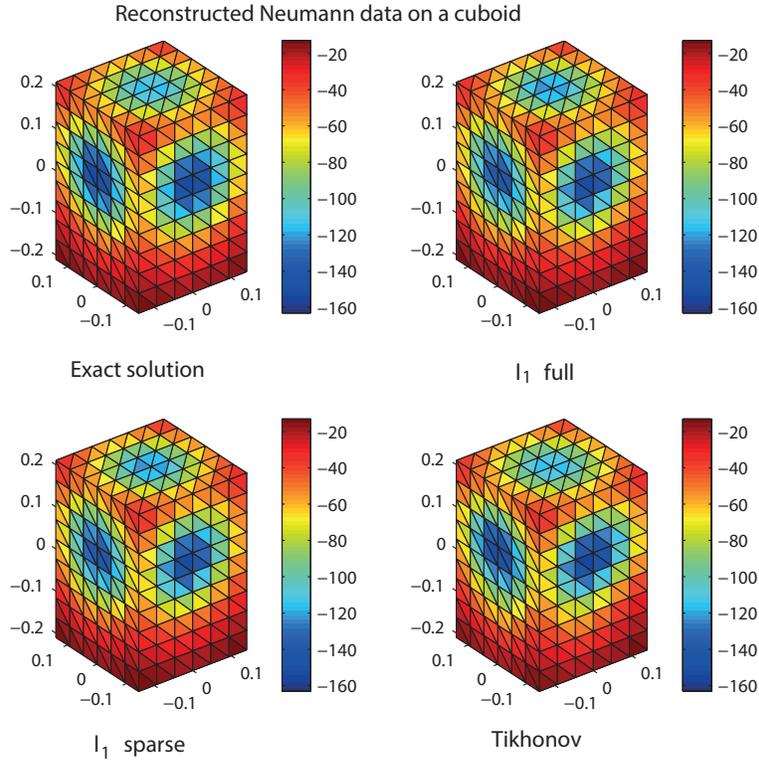, width=\textwidth}
        \caption{\label{fourplot} Neumann boundary data on a cuboid generated by a point source at $(0,0,0.05)$ with wavenumber $k=1$ and $15\%$ added noise. The plots compare the exact solution against those reconstructed using Tikhonov regularisation, the $\ell_1$ approach with all charge points and the sparse $\ell_1$ approach using only dominant charge points.}
\end{center}
\end{figure}
Figure \ref{azzplot} highlights the differing nature of the
solutions reconstructed using the Tikhonov and $\ell_1$ approaches.
The plots show that $\ell_1$ regularisation is more effective at
promoting sparsity in the case of noisy data, and hence larger
values of the data fidelity constraint $\epsilon$. In particular,
when noise is added the solutions can be accurately reconstructed
using only 10 to 15 of the 576 source points. This is further
illustrated in Figure \ref{fourplot} which shows the reconstructed
solution with noise level $w=0.15$ (or equivalently
$\mathrm{SNR}_{\mathrm{dB}}=16.48\mathrm{dB}$) using each of the
three solution strategies described above. The exact solution is
also shown for reference.

In all three cases we achieve a faithful reconstruction of the
Neumann data on the cuboid since the match to the exact solution is
very good. Plots of the cases $w=0$ and $w=0.05$ are omitted for
brevity, since as shown in Table \ref{T1}, the reconstruction errors
in these cases are even smaller and the likeness to the exact
solution shown in Figure \ref{fourplot} would be even stronger.
\emph{The main result of this section is that $\ell_1$ sparse
regularisation can give similar accuracy to Tikhonov regularisation
for noisy data, but with a small fraction of the number of charge
points required to produce the reconstruction.}

\subsection{Higher and irregular frequencies}

We now investigate the behavior of the method for some potentially
problematic choices of the wavenumber $k$. First we look at the case
when the frequency is increased, including when the Nyquist
frequency is exceeded. Since our measurements are taken at triangle
centroids then the resulting measurement grid is irregular and so
the Nyquist frequency is not well-defined. We therefore choose the
Nyquist frequency associated with the regular grid given by the
triangle vertices, as a value approximately representative of the
Nyquist frequency. For the discretisation considered in the previous
section with $576$ triangles, the grid spacing is $\Delta x =
0.04667$, meaning that the wavenumber corresponding to the Nyquist
frequency is $k_{\mathrm{nyq}}=\pi/\Delta x= 67.32$. We also
investigate the performance of the method close to other typical
threshold frequencies for numerical solution approaches, such as the
six grid points per wavelength rule of thumb for finite and boundary
element methods, which gives a maximum wavenumber of $k=22.44$ for
the grid described above. The performance of the method at irregular
frequencies will also be investigated. For the method of
superposition these irregular frequencies are the resonances of the
region enclosed by the interior source surface \cite{GK89}.
Numerical studies indicate that one such frequency is close to
$k^*=17.54/\alpha$, which here is $k=52.62$.

Table \ref{T2} gives the reconstruction error for a range of
wavenumbers $k$ using $\ell_1$ reconstruction techniques and
compares the accuracy of the reconstruction using all 576 charge
points, and using two different values of the sparsity parameter
$\beta$. In particular, we compare the default choice used in the
last section of $\beta=0.5$ with the choice $\beta=0.9$, which uses
fewer charge points but at the potential cost of poorer accuracy.
The maximum wavenumber studied corresponds to the wavelength being
close to (but still greater than) the exterior measurement distance
$\delta$. The results show that both irregular and high frequencies
lead to a degradation in the accuracy of the reconstruction, and
lead to a loss of sparsity in the reconstructions. Accurate and
reasonably sparse reconstructions can be generated provided there
are at least 3 data points per wavelength since for up to $k=44.88$
we can reconstruct the solution with a smaller error than the level
of added noise ($15\%$) and with at least an order of magnitude
reduction from the total number of charge points (576). These
results are consistent with the findings of Ref. \cite{AB08}, where
it is also suggested that a superposition method will give accurate
results provided there are at least 3 degrees of freedom per
wavelength. We note that if the surface of interior charge points
includes the monopole generating the acoustic field then one would
obtain exact representations for arbitrarily high frequencies.

\begin{figure}[h]
\begin{center}
          \epsfig{file=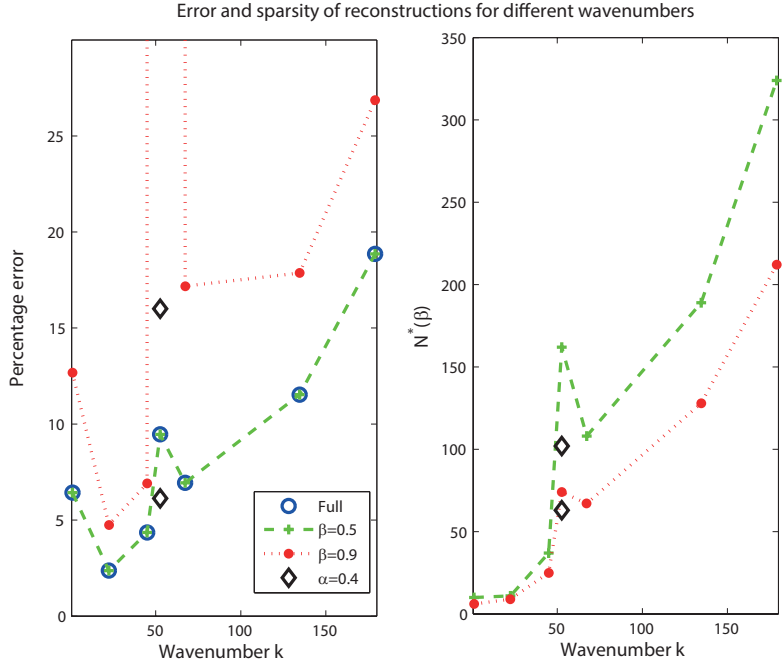, width=\textwidth}
        \caption{\label{kchange} The accuracy and sparsity of the reconstructed solutions with $15\%$ added noise, $\alpha=1/3$ and $z_{0}=0.05$. The plots show the effect of changing the wavenumber $k$, including the effect of irregular frequencies and values above the Nyquist frequency.}
\end{center}
\end{figure}

In addition to the general trend of increased errors for higher
frequencies, one also observes a local peak in the error at the
characteristic wavenumber $k=52.62$. Here we see that the error is
particularly poor for the sparse reconstruction with $\beta=0.9$,
and that there is also a local peak in the number of dominant charge
points. This suggests that sparse reconstructions are not feasible
at higher frequencies or at irregular frequencies. However, the
reconstruction error is lower than the noise level for the schemes
using all charge points or with $\beta=0.5$ for all frequencies
tested up to twice the Nyquist frequency. The results of this
section therefore suggest that the method of superposition with
$\ell_1$ regularisation can provide excellent reconstructions for
frequencies up to around twice the Nyquist frequency, and that
sparse reconstructions are feasible provided we have at least three
data points per wavelength. Irregular frequencies degrade both
accuracy and sparsity. However, if a more accurate and sparsely
reconstructed solution was required at $k=52.62$, then we could
change the scaling of the internal source surface (i.e. change
$\alpha$) which would move the location of the irregular frequency.
Changing $\alpha$ from $1/3$ to $0.4$ leads to a percentage error of
\textsc{6.131\%} for both the full reconstruction and the sparse
scheme with $\beta=0.5$, which identifies $N^*(0.5)=102$ dominant
charge points. For $\beta=0.9$, the error increases to
$\textsc{16.00\%}$ with $N^*(0.9)=63$. Note that these results are
far more consistent with the other results shown in Table \ref{T2}
and the removal of the local error peaks is shown more clearly in
Fig. \ref{kchange}, where the diamond symbols show the values
computed with $\alpha=0.4$.

The results presented by Chardon \emph{et. al.} \cite{GC12} using
sparse plane wave reconstructions indicate that randomising the
exterior data point locations (measurement locations) within the
hologram plane facilitates sparse reconstructions above the Nyquist
limit. Unfortunately, the reconstructed solutions using the method
of superposition lose their sparsity at frequencies around and above
the Nyquist limit and randomising the data point locations has been
observed to degrade the accuracy of the reconstructed solution in
general across the range of wavenumbers studied in Fig.
\ref{kchange}. These observations are consistent with the findings
in Ref. \cite{MB10}. \emph{The main result of this section is that
although in certain special cases  $\ell_1$ sparse regularisation
could give exact representations up to arbitrarily high frequencies,
in general the reconstruction accuracy will decrease at higher
frequencies. Irregular frequencies can also be treated simply by
perturbing the surface of interior charge points.}

\subsection{Dependance on the singularity and charge point locations}

The results in the previous sections all reconstructed the Neumann
data generated from a point source located on the $z-$axis at
$z_{0}=0.05$. This ensured that the singularity in the solution of
the related interior problem was located within the interior charge
point surface for $\alpha=1/3$. We now consider how the accuracy and
sparsity of our reconstructed solutions depends on both the position
of an interior source point generating the exterior pressure data,
and the relative size/ position of the internal charge point surface
controlled by the parameter $\alpha$ as described in Eq.
(\ref{alpha}).

\begin{figure}[h]
\begin{center}
          \epsfig{file=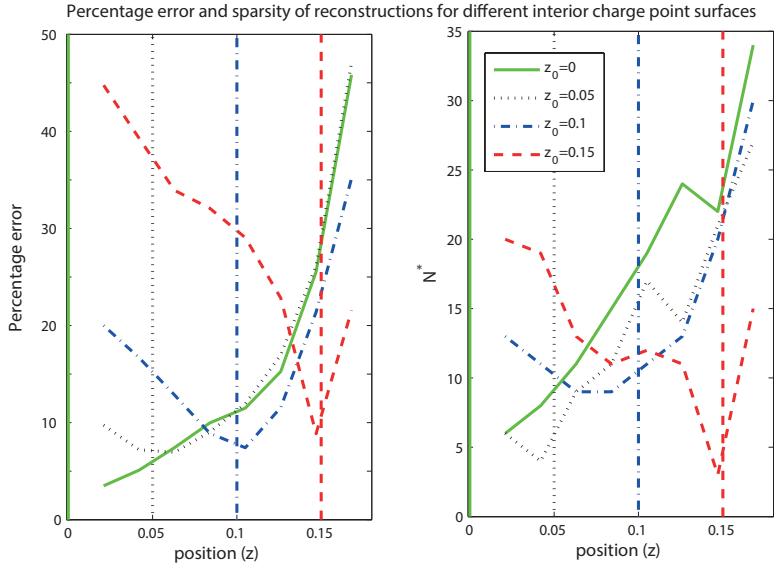, width=\textwidth}
        \caption{\label{zchange} The accuracy and sparsity of the reconstructed solutions with $k=1$ and $15\%$ added noise. The plots show the effect of using a range of different sized interior charge point surfaces and different positions for the source point generating the boundary data.}
\end{center}
\end{figure}

Fig. \ref{zchange} shows both the percentage errors for the sparse
reconstructions and the value of the sparsity parameter $N^*(0.5)$
for different sized interior charge point surfaces and for different
positions of the point source generating the exterior data. These
quantities have been computed for values of $\alpha$ between $0.1$
and $0.8$, and for $z_{0}$ between $z_{0}=0$ and $z_{0}=0.15$.
Instead of showing values of the parameter $\alpha$, Fig.
\ref{zchange} shows the corresponding $z-$coordinate where the
interior charge point surface intersects the positive $z-$axis. In
this way we are able to indicate the size of the interior charge
point surface relative to the location of $z_{0}$ on the same axes.
Note that since $\Gamma$ intersects the positive $z$-axis at
$z=0.21$ (it is centred at the origin with total height $0.42$m),
then the internal source surface with, for example, $\alpha=0.5$
will intersect the positive $z$-axis at $z=0.105$, and this is the
value used along the horizontal axis in Fig. \ref{zchange}. In all
cases the added noise level is $15\%$. Note that the relative errors
obtained when reconstructing the solution using all charge points
differs from that given by the sparse reconstruction by less than
$1\%$.

Each subplot of Fig. \ref{zchange} shows four curves, corresponding
to each of four choices of $z_{0}$, and four vertical lines
indicating the positions of the corresponding point $z_{0}$. The
left subplot shows the percentage errors for different choices of
interior source surface. We notice that the errors are minimised
when the size of the interior source surface is such that it
intersects the positive $z-$axis close to $z=z_{0}$, where the curve
crosses its corresponding vertical line. Likewise, this right
subplot shows that the number of charge points $N^*$ needed to
obtain a sparse reconstruction is also minimal when the size of the
interior source surface is such that it intersects the positive
$z-$axis close to $z=z_{0}$. In general, the solutions are
reasonably accurate for source surfaces intersecting the positive
$z-$axis between $z=0.05$ and $z=0.1$, corresponding to choosing
$\alpha=0.3$ or $\alpha=0.4$. Choosing $\alpha=0.8$ so that the
source surface intersects the positive $z-$axis at $z=0.168$ gave
the worst results in general. Interestingly, the results of this
section suggest that it doesn't seem to be critical whether or not
the surface of interior charge points encloses any singularities in
the modelled wave field. Furthermore, the results also point to
important potential applications of the sparse superposition method
developed in this work for source identification problems in
general.

\emph{The main result of this section is that both the accuracy and
sparsity of the reconstructed solution is enhanced when the surface
of interior charge points includes points close to the location of
the interior monopole generating the acoustic field. A major
strength of the sparse reconstruction approach described here will
therefore be in its application to source identification problems.}

\subsection{Example of a locally radiating structure}

In this section we consider the problem of reconstructing the
vibrations of a structure which is rigid, except over a relatively
small region. This is both a typical assumption for applications to
loudspeakers, and is also typical of problems commonly modelled
using the method of patch NAH, whereby measurements and
reconstructions only take place in the vicinity of the vibrating
region (see for example Refs. \cite{AS05,EW03}). In the present
study we use such an example for verification of the sparse
superposition method for a problem where the pressure data is not
generated by a monopole point source, and hence an optimal choice
for the surface of internal charge points would not be related to
the location of such a monopole.

\begin{figure}[h]
\begin{center}
          \epsfig{file=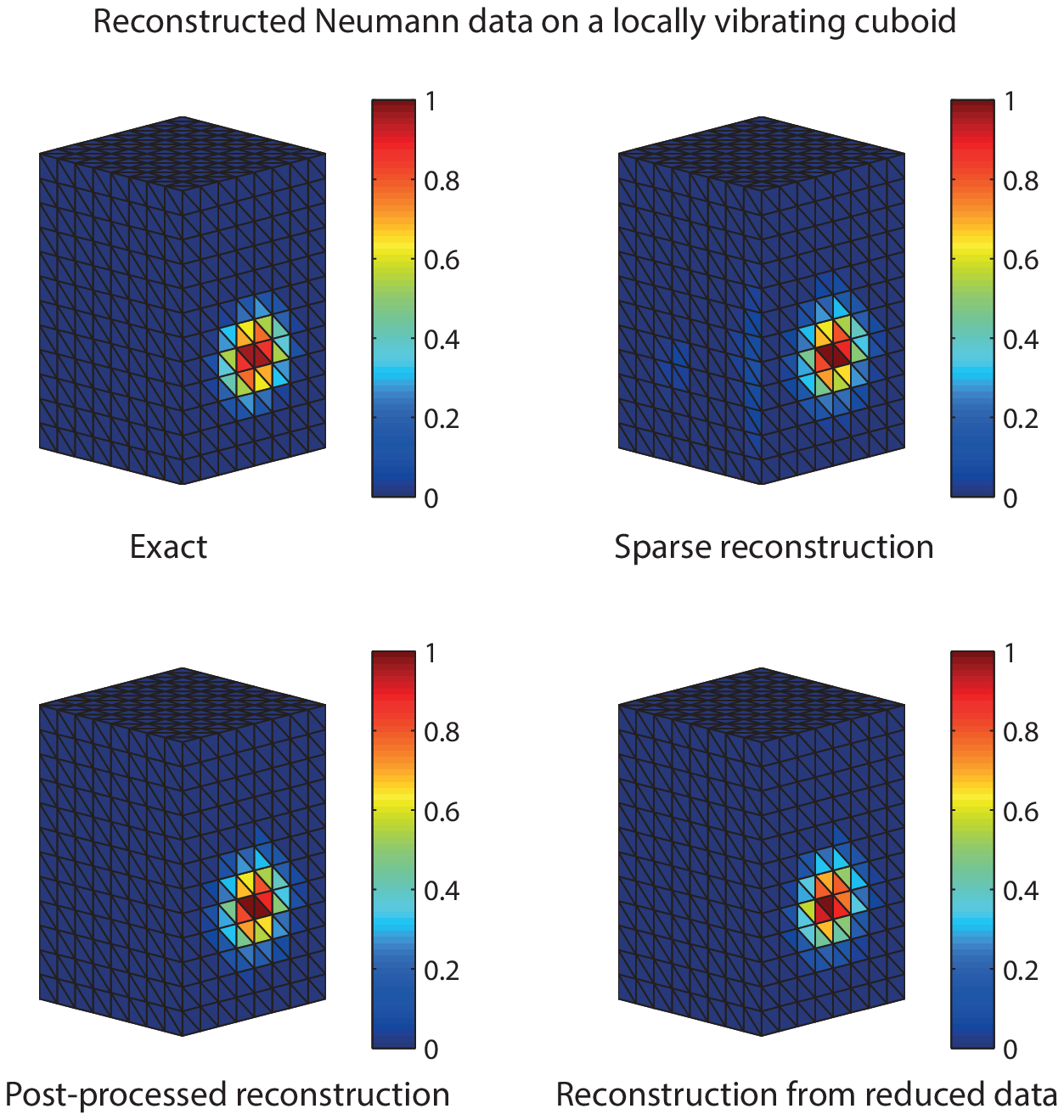, width=\textwidth}
        \caption{\label{twoplot} The Neumann boundary data for a locally radiating structure with $1024$ charge points, $1024$ data points, $k=1$ and $\alpha=1/3$. Comparison of the exact solution (upper left) with the sparsely reconstructed boundary data (upper right), the sparsely reconstructed boundary data after post-processing (lower left) and a reconstruction using only 28 charge points and a reduced data set with 800 values (lower right).}
\end{center}
\end{figure}

We consider reconstructing Neumann boundary data given by a raised
cosine function
\begin{equation}\label{bcloc}
\frac{\partial
p}{\partial\mathbf{n}_{x}}(\mathbf{x})=\frac{1}{2}\left(1+\cos(10\pi
|\mathbf{x}-\mathbf{x}_0|))\right),
\end{equation}
inside the circle defined by
$\Gamma\cap\{\mathbf{x}:10|\mathbf{x}-\mathbf{x}_0|<1\}$, with
$\mathbf{x}_0=(0.14,0,-0.0525)$. Outside of this circle we let the
Neumann boundary data be zero. The exterior pressure data at a
distance $\delta=0.035$m from $\Gamma$ are generated using a
Burton-Miller based BEM as described in Ref. \cite{SA92} for
example. The triangulation has been refined compared to the results
in previous sections and now has $1024$ triangles / interior source
points and $1024$ exterior data points. This has been done to
improve the resolution of both the boundary data representation and
the BEM approximation of the exterior pressure data. Note that here
there is an extra source of noise (in addition to the $15\%$ added
noise) in the acoustic pressure data due to the numerical error in
the BEM approximation. The results of the sparse reconstruction
technique for wavenumber $k=1$ are shown in Fig. \ref{twoplot}. We
found that a value of $\alpha$ between $0.3$ and $0.4$ gave the best
results, which is consistent with the previous section, and hence we
have taken $\alpha=1/3$.

The upper subplots of Fig. \ref{twoplot} show that the sparse
superposition method produces a good reconstruction of the vibrating
region for the prescribed locally vibrating boundary data. The
computations shown use $\beta=0.5$ to generate the sparse scheme
leading to $N^*=30$ dominant charge points (of $1024$ in total). The
percentage error in the sparse reconstruction is \textsc{21.19\%},
which is the same (up to the quoted level of accuracy) as the error
using all $1024$ charge points for the reconstruction. We note that
the boundary data to be reconstructed has $52$ out of $1024$ entries
that are non-zero. Relatively significant errors arise in the quiet
regions of the locally vibrating object, close to the edges of the
cuboid that are nearest the vibrating region. If we assume prior
knowledge of the non-vibrating regions and consider only the
accuracy of the reconstruction over the vibrating region, then the
error reduces to \textsc{9.843\%} for both the sparse and full data
reconstructions. A plot of this post-processed result is shown in
the lower left subplot of Fig. \ref{twoplot}.

The lower right subplot of Fig. \ref{twoplot} shows the result of
reconstructing the Neumann boundary data starting only with the $30$
dominant source points identified by the sparse reconstruction
method and using a reduced data set sampled from the full data set
at $800$ randomly selected points. The same post-processing
procedure as described above has been applied to fix the solution in
the non-vibrating regions to zero. The results shown are computed
using Tikhonov regularisation (with only the 30 charge points
identified by the initial sparse reconstruction), which gave
slightly better accuracy than using the sparse reconstruction
algorithm for a second time. The percentage error in the plotted
reconstruction is \textsc{18.89\%}, compared with \textsc{26.36\%}
for the reconstruction without post-processing. Using the sparse
algorithm to reconstruct the solution instead gave errors around
$3\%$ higher in each case. Note that these results all show an
improvement on the reconstruction obtained from the same reduced
acoustic data set, but using a basis with all $1024$ charge points.
In this case the percentage error was more than doubled to
\textsc{58.87\%}, and improved to \textsc{46.52\%} after
post-processing. We remark that the reconstruction using the full
basis with reduced data is an under-determined problem (the acoustic
data are fewer than the number of unknowns), whereas the
reconstruction from the sparse basis is over-determined. This
suggests that reducing the number of charge points and changing the
under-determined problem into an over-determined one is an important
step for the efficient implementation of NAH with reduced data using
the method of superposition. \emph{The main result of this section
is that sparse reconstruction methods can still be applied when a
suitable choice of dictionary of basis functions for the sparse
reconstruction is not obvious, provided there is some inherent
sparsity that can be exploited. Sparse reconstructions can also be
used in conjunction with reduced acoustic field data sets giving
reasonable results. In particular, the sparse basis representation
leads to better accuracy and more efficient calculations than using
the full basis with the reduced data set.}

\section{Conclusions}

The method of superposition has been combined with a sparse $\ell_1$
reconstruction algorithm and applied to the problem of near-field
acoustic holography. The developed sparse superposition method is
able to reconstruct the normal velocity of a vibrating object using
only a very small number of charge points in many cases, in contrast
with a standard Tikhonov reconstruction. In particular, it appears
that competitive sparse reconstructions can be generated provided
the wavenumber is not too large and that the data can be assumed to
be sufficiently noisy to permit a data fidelity parameter of at
least $5\%$ of the size of the $\ell_2$ norm of the exterior
pressure data. Sparsity also appears to be an important factor when
considering reduced acoustic field data sets, where reconstructions
using a sparse basis gave a considerable improvement in accuracy
over using the full basis with all charge points. The results of the
simulations point to a number of important future developments. The
sparse superposition method could be used to generate initial
conditions for the method of fundamental solutions, whereby the
accuracy of the reconstruction is optimised over both the charge
point strengths and locations using a non-linear optimisation
algorithm. This technique could also be applied more widely to
source identification problems, as the results have shown that when
the charge points are close to point sources generating the exterior
pressure data then the reconstruction becomes very sparse, and the
charge points close to the generating point source are highly
dominant.

\section*{Acknowledgement}
N.M. Abusag gratefully acknowledges financial support from the
Libyan Ministry of Higher Education and Scientific Research. D.J.
Chappell gratefully acknowledges financial support from the European
Union (FP7-PEOPLE-2013-IAPP grant no. 612237 (MHiVec)). We also
thank Prof. Carl Brown for carefully reading the manuscript.

\begin{table}[th]
\caption{\label{T1} The $\ell_2$ percentage error in the
reconstructed Neumann boundary data generated from a source point on
the $z-$axis at $z_{0}=0.05$ with added noise and $k=1$.}
{\begin{tabular}{@{}cccc@{}} \hline\\ {Noise level $(\%)$} & {\%
error for $\ell_1$ full } & {\% error for $\ell_1$ sparse } & {\%
error for Tikhonov}
\\
\hline\\
0  & \textsc{6.2091e-3}   & \textsc{0.1097}  & \textsc{1.009e-7}  \\
  5  & \textsc{3.850}   & \textsc{3.850} & \textsc{3.549} \\
  15 & \textsc{6.428}   & \textsc{6.429}  & \textsc{6.017}  \\
\hline
\end{tabular}}
\end{table}

\begin{table}[th]
\caption{\label{T2} The $\ell_2$ percentage error in the
reconstructed Neumann boundary data generated from a source point on
the $z-$axis at $z_{0}=0.05$, with internal source surface at
$\alpha=1/3$ and $15\%$ added noise over a range of wavenumbers
$k$.} {\begin{tabular}{@{}cccccc@{}} \hline\\
{$k$} & {\% error: full} & {\% error: $\beta=0.5$ } & {$N^*(0.5)$} & {\% error: $\beta=0.9$ } & {$N^*(0.9)$}\\
\hline\\
1  &     \textsc{6.428}   & \textsc{6.429}  & 10 & \textsc{12.27}  & 6  \\
  22.44  & \textsc{2.369}   & \textsc{2.369}  & 11 & \textsc{4.752}  & 9  \\
  44.88  & \textsc{4.358}   & \textsc{4.358}  & 37 & \textsc{6.904}  & 25 \\
  52.62  & \textsc{9.458}   & \textsc{9.462}  & 162 & \textsc{4797}  & 74 \\
  67.32     & \textsc{6.937}   & \textsc{6.937} & 108 & \textsc{17.18}  & 67  \\
  134.6     & \textsc{11.15}   & \textsc{11.15} & 189 & \textsc{17.86}  & 128 \\
  179    & \textsc{18.89}   & \textsc{18.89} & 324 & \textsc{26.87}  & 212 \\
\hline
\end{tabular}}
\end{table}
\end{document}